\documentclass[reprint,aps,prl,twocolumn,superscriptaddress]{revtex4-1}
\usepackage{graphicx}
\usepackage{amssymb,amsmath}
\usepackage{epstopdf}
\usepackage{slashed} 
\newcommand{\f}{\begin{equation}}
\newcommand{\ff}{\end{equation}}

\begin{document}

\title{The Equivalence Principle and the Emergence of Flat Rotation Curves}


\author{Stephon Alexander}
\affiliation{Department of Physics, Brown University, Providence, RI, 02906}
\author{Lee Smolin}
\email{lsmolin@perimeterinstitute.ca}
\affiliation{Perimeter Institute for Theoretical Physics, 31 Caroline Street North, Waterloo, Ontario N2J 2Y5, Canada}
\date{\today}
\begin{abstract}
We explain flat rotation curves and the baryonic Tully-Fisher relation by a combination of three hypotheses.  The first is a formulation of the equivalence principle for gravitationally bound quantum $N$ body systems,  while the second is a second order phase transition hypothesized to arise from a competition between the effects of Unruh and deSitter radiation experienced by a static observer in a galaxy.  The third is a light dark matter particle, coupled to a dark photon.  

The phase transition is triggered in a ring where  the Unruh temperature of a static observer falls below the deSitter temperature, thus explaining the
apparent coincidence that Milgrom's $a_0 \approx a_\Lambda = c^2 \sqrt{\frac{\Lambda}{3}}$
  This phase transition drives the dark matter particles to a regime characterized
by a broken $U(1)$ invariance and an approximate scale invariance.  In this regime, the dark matter condenses to a supercurrent characterized by a differentially rotating ring with a flat rotation curve, coupled to a dark magnetic field.   The baryonic Tully Fisher relation is a direct consequence of the
approximate scale invariance\cite{MM-scale}.

\end{abstract}
\maketitle



\section{Introduction}

The standard $\Lambda CDM$ paradigm of collisionless cold dark matter is successful at explaining the missing mass problem\cite{Vera} on large scales.  However, there are a handful of discrepancies on galactic scales which motivate 
a modification of the $\Lambda CDM$ paradigm. Of particular interest is the observed baryonic Tully-Fisher relation\cite{Moti1} which reveals a tight coupling between the disk galaxy's asymptotic rotational velocity and its total baryonic mass, $v_{c}^{4} = Ga_0 M_{b}$, where 
$a_0 = 1.2 \times 10^{-10} m/s^2$ is a critical acceleration, read off the 
data\cite{TF,small-scatter,McGaugh:2000sr} and $M_b$ is the galaxy's baryonic mass. There is also observed a relatively tight relation between the acceleration a star would have due to Newtonian gravity and baryonic matter and the actual observed accleration\cite{Milgrom,MLS}.  In addition, there is the fact that $a_0 $ is remarkably close to the acceleration of the universe due to dark energy, 
\f
a_\Lambda = c^2 \sqrt{\frac{\Lambda}{3}}.  
\label{aL}
\ff
This intriguing fact, which may be stated as saying
that the acceleration of the universe gives a typical scale for the acceleration of stars in galaxies, is not explained by $\Lambda CDM$, and was an early motivator of the MOND 
paradigm\cite{MM-related}.

Collisionless cold dark matter does not account for the baryonic Tully-Fisher relation, the acceleration relation,  and several  small scale anomalies, such as the core-cusp and missing satellite problems.  Modified newtonian gravity\cite{Milgrom,MOND-review,MM-related} (MOND) does account well for these data, but is challenging to extend to a stable relativistic theory\cite{teves,moffat} 
and fails to account for the evidence for dark matter in clusters, lensing and the $CMB$. Recently there have been attempts to build MOND-like behaviour into the physics of dark matter\cite{minic}, including the intriguing suggest that dark matter has a 
superfluid phase\cite{Silverman}-\cite{Alexander2}.  However so far these inspiring suggestions do not completely account for the
closeness of $a_0$ to $a_\Lambda$
We build on these ideas here.

This leads us to propose a new scenario, which develops the idea that dark matter has a  super-fluid phase, which, if successful, would explain the flattening of the rotation curves, the Tully-Fisher and acceleration relations and the relation between $a_0$ to $a_\Lambda$.   This is based on three hypotheses. The first two are related to how the equivalence principle may be realized in the quantum domain\cite{Smolin}.

\begin{enumerate}

\item{}  Consider a non-relativistic gravitationally bound  quantum $N$ body system, with identical masses and only gravitational interactions.   Because of the non-locality of quantum effects, the standard formulations of the equivalence principle are broken by effects proportional to $\hbar $.  To begin with there are three roles for mass, the passive and active gravitational masses, $m_p$ and $m_a$ and the inertial mass, $m_i$. 

{\it The weak, non-relativistic quantum equivalence principle} posits that
in the limit of large $N$, these quantities only occur in two combinations.
One with $G$  (which is the only place $G$ occurs) 
\f
Q= Gm_a
\label{Q}
\ff
measures the strength of the active gravitational force, and
the {\it quantum diffusion constant}
\f
D= \frac{\hbar}{m_i} =  \frac{\hbar}{m_p}
\label{D}
\ff
In particular the semiclassical limit $\hbar \rightarrow 0$ is defined by
\f
D \rightarrow 0
\ff
in which limit $m_i$ and $m_p$ disappear. 

\item{}{\it The thermal equivalence principle} The physics of a quantum $N$ body system which generates a bulk static or stationary gravitational field is to be described by static observers, who exerts an acceleration $a$ to stay static\cite{Smolin}.  The spacetime as a whole may also experience an acceleration, 
$a_\Lambda$ due to dark energy or a cosmological constant.

Then the static observer experiences a coupling to two different heat baths at 
different temperatures
\f
T_a = \frac{\hbar a}{2 \pi c}, \ \ \ \mbox{and} \ \ T_\Lambda = \frac{\hbar a_\Lambda}{2 \pi c}.
\ff

\item{}We hypothesize a very light dark matter particle coupled to a dark electromagnetic-like 
field.

\end{enumerate}

Our basic hypotheses then are that 

\begin{itemize}

\item{}The dark matter is extremely light and so forms a quantum fluid.  (Models of dark matter superfluidity can be found in \cite{Silverman}-\cite{Alexander2}).  This requires that
\f
\lambda_{dB} = \frac{\hbar}{m_i v} > \rho^{-\frac{1}{3}}
\label{criteria}
\ff
\item{}Competition between the effects of $T_a$ and $T_\Lambda$
drives a second order phase transition whose order parameter is the velocity field of the quantum fluid, so the cold phase, for $T_a < T_\Lambda$, is a superfluid.

\item{}The second order phase transition enforces a scale invariance
on the superfluid phase. As Milgrom has pointed out in \cite{MM-scale} scale invariance is nearly a sufficient condition to recover MOND-like behavior,  including a tendency for the rotation curves of the dark matter particles to flatten when their accelerations are near the critical value of $a_\Lambda$.
The scale invariance also explains the baryonic Tully-Fisher relation\cite{MM-scale}.

Note that the scale invariance is imposed only within a ring
$r_0 < r < r_2 $ in which accelerations are sufficiently near the critical point that scale
invariant behaviour may be expected.  We call this ring the critical region. 

\item{}For  simplicity we assume that most of the baryonic matter lies within the inner
boundary $r_0$, and that the dark matter disk is confined to the  ring. 

\item{}The superfluid requires coupling to a dark magnetic field to have stable rotating solutions, with constant velocity, to the semiclassical GP equation, making for a dark ring of super-currents.

\end{itemize}

Thus, within the critical region, there is an emergent quantum phenomenon that organizes the dark matter super-currents,
so that it obeys approximate 
scale invariance, and a consequence is a flattening of the  rotation curves for the stars as
well as for the dark super-current..

Hence, this scenario explains the acceleration relation, the baryonic Tully Fisher relation, and the closeness of
$a_0$ to $a_\Lambda$, as well as the fact that the
MOND-like behaviour is only seen for a limited regime and does not extend to arbitrarily small acceleration.

In this letter we give a  brief sketch of this new scenario, many details remain to be worked out.

We proceed in three steps.  First we discuss the physics of a charged superfluid when scale invariance
is imposed on it in a ring shaped region.  Second, we discuss the motions of stars within the gravitational field created by the baryons in the core and  the dark superfluid ring.
Third, we discuss a hypothesis as to the origin of
that scale invariance.

\section{The dark matter as a quantum fluid}

We consider a gravitationally bound population (or sub-population) of dark matter particles which we will assume has condensed into a thick disk.  The hypothesis that a portion of the dark matter condenses to form a disk has been argued via. a few mechanisms, including dissipative, self interactions and mergers\cite{Reade,Spergel,JiJi,Saavas}.  Thin disks have been ruled out by observation from Gaia but thick disks still remain a possibility\cite{thick}. In this work we do
not enquire into the mechanism of disk formation, nor do we discuss what proportion goes into the disk and which  stays in the halo; we simply assume that a dark disk has formed. 
(However we might mention that coupling to a dark photon could play a role in formation of a disk\cite{JiJi}).
We note that the arguments we make below would still be valid were there a small self-interaction.  

We assume the particles are spinless bosons of mass $m$, which is chosen so that the 
criteria (\ref{criteria}) to be described as a quantum fluid is satisfied.


We describe the population by a complex many body wave-function,
$\Psi (\{ x^a_I, t \} )$, whose dynamics is specified by a non-relativistic Schrodinger equation.
\f
\imath \hbar \frac{\partial \Psi}{\partial t} = \left \{ -\sum_K  \frac{\hbar^2}{2m_{i}} \nabla_K^2 
- \sum_{K< J} \frac{G m_{a }  m_{p }}{|x_J-x_K |}  
\right \} \Psi
\label{Sch1}
\ff
where $\nabla^2 = \frac{1}{\sqrt{q}}\nabla_a \sqrt{q}q^{ab} \nabla_b $, with $q^{ab}$ the flat metric and $\nabla_a$ is a $U(1)$ covariant derivative, providing coupling to a dark $U(1)$
gauge field.
\f
\nabla_a \Psi = \left (
\partial_a + \imath A_a
\right ) \Psi
\ff
The $U(1)$ gauge invariance  is
\f
A_a \rightarrow A^\prime_a = A_a - \partial_a \xi, \ \ \ \ \ \Psi \rightarrow \Psi^\prime
= \Psi e^{\imath \xi }
\ff

There is also the gravitational self-interaction.
In addition, our argument will not be much altered if there is a small self-interaction, with small dimensionless coupling $\lambda$.  We also do not discuss the implications of the dark 
Maxwell equations which determine the configuration of the dark magnetic field off the disk, save for a speculation mentioned at the end.

Note that we can divide through by the passive gravitational mass $m_{p}$, to find that the resulting equation depends on the inertial or passive gravitational mass only indirectly, through the quantum diffusion constant defined by (\ref{D} )
through
\f
\imath  {D} \frac{\partial \Psi}{\partial t} = \left \{ -  \frac{ {D}^2}{2} \sum_I \nabla_I^2 
- Q \sum_{I< J} \frac{1}{|x_J-x_I |}
\right \} \Psi
\label{SchD2}
\ff

The  passive and inertial masses are absorbed into $D$, which is to say $\hbar$, and the only place the active mass $m_a$ appears is in the bulk potential energy, where it is multiplied
by $G$ in the combination (\ref{Q}), which is the only place $G$ appears.  We now consider the classical limit, which is the limit $\hbar \rightarrow 0$.  But
note that the Schrodinger equation also only depends on $\hbar$ through its
dependence in ${D}$.  So the $\hbar \rightarrow 0 $ limit must be a 
${D} \rightarrow 0$ limit.  In this limit the passive gravitational mass and inertial mass go away.
This is consistent with the weak, non-relativistic quantum equivalence principle.


To get to the classical limit we write 
\f
\Psi [\{ x_I^a \}, t ] = \sqrt{\rho} e^{\frac{1}{\hbar} S}
\ff
where $S$ satisfies the Hamilton-Jacobi equation.   Now note that
\f
p_a = m_i \dot{x}^a_I = \nabla_a S
\ff
So we are interested in the specific action
\f
s ( \{ x_K \} , t ) =\frac{S ( \{ x_K \} , t )}{m_i}
\label{specificS}
\ff
which satisfies
\f
\dot{x}^a_I = \nabla_a s
\label{xdot}
\ff
This satisfies
\f
\dot{s}( \{ x_K \} , t )=  \sum_I  q^{ab} \nabla^I_a s \nabla^I_b s  - \sum_{I >J}  
\frac{G {m_a}}{|x-x_I |} + \frac{1}{8} {D}^2 \frac{\nabla^2 \sqrt{\rho}}{\sqrt{\rho}}
\label{HJ1}
\ff
Now we take the limit ${D} \rightarrow 0$ and all this does is remove the last, quantum
potential term, leaving us with the classical Hamilton Jacobi equation for the specific action
\f
\dot{s}( \{ x_K \} , t )=  \sum_I  \frac{1}{2} q^{ab} \partial^I_a s \partial^I_b s  +V
\ff
where the specific gravitational potential energy is
\f
V= - \sum_{I >J}  \frac{G {m_a}}{|x_I -x_J |} 
\ff

Note that the dependence on the masses  is now restricted to the total
gravitational potential, i.e. it involves only the active gravitational mass, $G m_a$.  This means that the trajectories given by (\ref{xdot}) are independent of the masses.This is a statement of the equivalence principle, discovered in the classical limit of a quantum self-gravitating system.

We should then
require that in the same limit, the wave function is symmetric amongst these identical particles.
\f
\Psi (x^a_I, x^a_J,\ldots ,t ) = \Psi (x^a_J, x^a_I ,\ldots ,t )
\ff
This implies
that the probability distribution $\rho (x^a_I )$ should be unchanged under an
interchange of particles, so we have
\f
\rho (x^a_I, x^a_J,\ldots  ) = \rho (x^a_J, x^a_I ,\ldots  )
\ff

We expect the same of the specfic Hamilton-Jacobi function of the averaged system
\f
s (x^a_I, x^a_J,\ldots ,t ) = s (x^a_J, x^a_I ,\ldots ,t )
\ff


\section{Physics of the quantum liquid}

The conditions we have found define a new phase of the quantum system of
$N$ bodies gravitationally bound via Newtonian gravity.  The system is in a semiclassical
regime, but it is a condensed system in which the symmetrization of the wave function is important.  We will see that this is because the symmetrization can impose on the bodies a super-fluid like behaviour. 

In the classical limit we have a specific Hamilton-Jacobi functional
(\ref{specificS}),
which is symmetric under exchanges of the particle labels.  
The velocities are given by (\ref{xdot}).

We  will impose three conditions on the wave function.
The first two are,

\begin{itemize}

\item{}{\bf Independence}  The particles are non-interacting, apart from their contributions
to the bulk gravitational potential, which is already separated out in the mean field approximation.
\f
s(\{  x^a_I \}  ) = \sum_I s_I(x^a_I )
\ff

\item{}{\bf Identical particles, condensed into a single state}, so that 
\f
s_I(x^a_I ) =s (x^a_I )
\ff

\end{itemize}


Given these two conditions, we can express the many body physics in terms of a single particle-like macroscopic wave function,
\f
\Phi (x^a , t ) =\sqrt{\rho (x)} e^{\imath \phi (x,t)}
\ff
where the density $\rho ( x,t)$ is defined from the many particle wave function 
and $\phi$ is the common global phase.  $\Phi$ satisfies the normalization
\f
N = \int d^3 x \rho (x; t ) 
\ff

In this mean field approximation we  can represent the gravitational potential
energy by its mean field value in terms of $M(r)$
\f
V(r) = - \frac{G  M(r) m }{r} 
\label{VM2}
\ff
where $M(r)$ is a functional of the density,
$\rho (x)$.

 In cylindrical coordinates $(r,\theta,  z)$,
\f
M[\rho ] (r) = 
2 \pi h m \int_0^r dr^\prime  r^{\prime } \bar{\Phi}(r) \Phi (r)
\label{M}
\ff
where $h$ is the height of the disk (assumed small).


The dynamics of $\Phi (x,t)$ are given by the self-consistent solution to
the non-linear Schroedinger equation
\f
\imath \hbar \frac{\partial \Phi}{\partial t} = \left \{ -\sum_I  \frac{\hbar^2}{2{m_i}} \nabla^2 
+V [\rho ](r) + {W_T}
\right \} \Phi
\ff

Here $W_T$ comes from varying the thermal contribution to the effective action, (\ref{VT})
which we will discuss below, but ignore for the time being.

This gives a version of the Gross-Pitaevskii (GP)  equation\cite{GP} (also called the Newton-Schroedinger equation.) 

As before, we divide through by $m_p$ to express the principle that the inertial and passive
gravitational masses and $\hbar$ are only expressed together in $D$.

\f
\imath D \frac{\partial \Phi}{\partial t} = \left \{ -\sum_I  \frac{D^2}{2} \nabla^2 
- {\cal V}  (\bar{\Phi}\Phi ) + {{\cal W}_T}
\right \} \Phi
\label{GP}
\ff

where the specific potential energy density is 
\f
{\cal V }= \frac{V}{m}
\label{VM3}
\ff

\subsection{A ring of dark super-current}

We now restrict our considerations to a ring defined by
\f
r_0 < r < r_2
\ff
We will later find it useful to assume that most of the baryonic mass is contained within an inner region for $r< r_0$.  We will impose scale invariance within the ring.  
We assume that the height of the disk $h << r_0$ so that vertical distributions can be ignored.

We assume that the wavefunction is of the form: 
$\Psi(\rho,z,\theta) =\rho (r)e^{i\phi (r, \theta , t)}$ 
where the z dependence is ignorable.  



We want to solve the real and imaginary parts of (\ref{GP}), which are the 
extended
Hamilton Jacobi
equation
\f
\dot{\phi}( r,\theta , t )=  \frac{1}{\hbar} E= \frac{1}{2} q^{ab} \nabla_a \phi  \nabla_b \phi  - 
{\cal V} (\rho ) + \frac{1}{8} {D}^2 \frac{\nabla^2 \sqrt{\rho}}{\sqrt{\rho}}
\label{HJ2}
\ff
and the current conservation law
\f
\dot{\rho} = - \nabla_a J^a
\label{prob}
\ff
where the  current is
\f
J_a = -\imath \bar{\Phi} \nabla_a \Phi + h.c. =    D \rho \left  ( \partial_a \phi + A_a \right )
=D \rho V_a
\ff

We also define the macroscopic phase $\phi$ by  a specific action
\f
\phi (x^a , t) = \frac{1}{D}(  \Sigma (x^a ) +e t )
\ff
where $e =\frac{E}{m}$ is the specific energy.

\subsection{Imposing scale invariance}

We now impose our third condition, which is limited to the ring
 $r_0 \leq r \leq r_2 $.  In this ring we impose scale invariance of the macroscopic wave function under scaling transformations defined by
 Milgrom \cite{MM-scale})
\f
t \rightarrow \lambda t , \ \ \ \ \ x^a \rightarrow \lambda x^a , \ \ \
r \rightarrow \lambda r,  \ \ \ \theta \rightarrow \theta
\label{scaling}
\ff
Physical constants such as $G$ and $\hbar$ don't scale, but masses do so we have
\f
m \rightarrow \lambda^{-1} m , \ \ \ \ \  D \rightarrow \lambda D
\label{scaling2}
\ff

\subsection{Using scale invariance to solve the GP equations}

We deduce from (\ref{scaling2}) that
\f
\Sigma (r, \theta) \rightarrow \lambda \Sigma (r, \theta) = \Sigma (\lambda r, \theta) 
\ff
from which we deduce that
\f
\Sigma = r f(\theta ) v
\ff
where $v$ is a constant speed.

We find solutions for
\f
\phi = \frac{1}{D} r \theta v, \ \ \ \ A_r =-\theta v, \ \ \ \ A_\theta =0
\label{solution}
\ff

Single-valuedness of the wave-function can be satisfied if the super-current 
is broken up into many discrete rings, which flow around circles at radii
\f
r_n = n \frac{D}{v} = n \lambda_{dB} 
\ff
which are integer multiples of the deBrogli wavelength.  However given (\ref{criteria}),  
we can expect
a large number of these rings fit within the critical ring, so we may ignore this microscopic structure
when considering the effect on the stellar orbits.

The current $J_a$ in  each ring is purely circular
\f
J_\theta = \rho rv , \ \ \ \ \ J_r=0
\ff
The norm of the velocity current, $V^a$,  is a constant
\f
||V||^2 = q^{ab}V_a V_b = v^2
\ff
which corresponds to a flat rotation curve.


It is not hard to show that to zeroth order in $D$ that, given $\phi$ of the 
form imposed by scale invariance (\ref{solution}),
 we can choose a 
form for the density, $\rho$, and a value of $E$, that together solve the GP equations,
in the form of (\ref{HJ2},\ref{prob}), 
to zeroth order in $D$.  

To solve (\ref{HJ2}) we need a constant ${\cal V}$, given by (\ref{M}) and (\ref{VM3}).
This self-consistent solution is given by
\f
\rho = \frac{v^2 }{2 \pi G r h}
\label{rho1}
\ff

To see this, we use, to zeroth order in $D$, the classical expression
\f
\frac{v^2}{r}= \frac{G M(r)}{r^2}
\ff
so
\f
M(r)= \frac{v^2 r}{G}= 2 \pi h {m} \int_0^r dr^\prime  r^{\prime } \rho (r') 
\label{M5}
\ff
which implies (\ref{rho1}).

(\ref{solution}) and (\ref{rho1}) also imply
\f
\dot{\rho} = 0 = \partial_a J^a
\ff
and so solve (\ref{prob}).

Meanwhile, the dark magnetic field is perpendicular to the ring and constant
\f
B^z = v
\ff
There is then an energy density in the ring from the magnetic field
\f
\rho_B = \frac{1}{2g^2} B^2 =  \frac{1}{2g^2} v^2
\ff
where $g$ is the coupling constant of the dark magnetic field.

Finally, we solve (\ref{HJ2}) to find that (neglecting the energy in the magnetic field))
\f
e =-\frac{{v}^2}{2} 
\ff
so that the viral theorem is satisfied.

The observation that the dark matter superfluid has a circular component in its motion is supported by 
the work of Reade et. al \cite{Reade}, which proposes that the history of accretion of satellites into the glactictic disc dynamically accretes a disc of dark matter.  As in our case,
this compoment of the dark disc is co-rotating with the visible disc.

This gives rise to a ring of  super currents, rotating around the galaxy for 
$r_0 < r < r_2 $.  

\subsubsection{Recovering baryonic Tully-Fisher}


We now deduce the orbital velocities of stars within the critical  ring.
At the inner boundary of the ring of the condensate phase, at radius
$r_{0}$,  we match a velocity 
${w}$ to the acceleration
\f
a_0 = \frac{{w}^2}{r_0}
\ff
Note that this will be the velocity of a star at the boundary.  It is not necessarily equal
to the velocity $v$ of the  dark super current because that is influenced by the dark 
magnetic field as well as the gravitational field. However we will show that
given our assumptions, the baryonic rotation curve is also flat within the ring.

Scale invariance dictates that in the outer region, between $r_0$ and $r_2$ the acceleration $a$
must fall off as
\f
a (r) = \frac{{w}^2}{r} 
\ff
where ${v}^2$ is constant (and hence scale invariant).  This is consistent with the dark
matter density of the form (\ref{rho1}), so that in the critical ring $M(r)$ grows proportionally in $r$
as our solution shows in (\ref{M5}).  Hence we find  the stars also have a 
flat rotation curve, given by $w$.

Assuming, then  that $M_b=M(r_0 )$ is the bulk 
of the baryonic mass, i.e. that most of the baryonic mass is in the inner region,
we can match Newton's law at $r_0$, by imposing $a_0=\frac{GM_b }{r_0^2}$ to
find
\f
r_0 =\sqrt{\frac{GM_b}{a_0}}
\ff
and the  baryonic Tully-Fisher relation,
\f
{w}^4 = GM_b a_0
\ff

So the flat rotation curves and the  baryonic Tully-Fisher relation are both consequences of the quantum equivalence  principle, 
and the non-relativistic equivalence principle, together with scale invariance.   

We note that Milgrom  \cite{MM-scale} emphasized that MONDian 
physics is characterized by scale invariance under (\ref{scaling}).

\section{Thermal physics of the global wave-function}

We have shown how scale invariance in the ring leads to a differentially rotating super-current with a constant current velocity, that
explains both the flat rotation curves and the baryonic Tully-Fisher relation.
Now we put forward a hypothesis to explain the origin of that scale invariance.

We consider the effects of the thermal accelerations.   We 
use Landau-Ginzburg theory to formulate the thermal physics in the mean field approximation.
But we also 
use the quantum equivalence
principle as formulated in \cite{Smolin} to organize the attribution of thermal effects to
accelerating systems.  This requires that, in static or stationary spacetimes, we associate a temperature to the accelerations experienced by static observers.  In our context these are non-rotating observers who employ a constant acceleration (say from a rocket engine) to maintain themselves at a fixed distance from the galactic centre.

Normally, when a quantum field is subject to a finite temperature $T$ we add a term to the effective potential energy
\f
V^T=  \frac{\alpha}{m} T^2   |\Phi |^2  
\ff
where $\alpha$ is a dimensionless combination of coupling constants.

Now, we assume that the dark matter (including the dark photon) couples to ordinary matter only through the gravitational interaction, thus it does not feel the $CMB$ and will not thermalize to its temperature.  It however does experience the universal thermal effects due to acceleration, with respect to inertial frames, whether with respect to local observers or cosmological horizons, as these are universal effects which
arise from the choice of vacuum.

We then have two sources of 
vacuum thermal 
effects\cite{Milgrom2}; these are 
the cosmological temperature\cite{dS} 
\f
T_\Lambda = \frac{\hbar a_\Lambda}{2\pi c} , \ \ \mbox{where} \ \ a_\Lambda = c^2 \sqrt{\frac{\Lambda}{3}}
\ff
due to radiation from the cosmological horizon and the Unruh temperature\cite{Unruh} due to the fact that the static observers are accelerating in the static 
gravitational field of the galaxy.
\f
T_a= \frac{\hbar a }{2\pi c}
\ff
Here we should emphasize that the physical situation we are describing is different from 
that experienced by a static observer in empty deSitter spacetime.  There the observe experiences the effects of being in equilibrium  with a single thermal bath, with the Deser-Levin temperature\cite{DL}
\f
T_{DL}= \sqrt{T_\Lambda^2 + T_a^2}
\ff
Here we have, instead, an observer who is out of equilibrium because she 
is coupled to two thermal baths, which have different temperatures,
one coming from  the cosmological horizon, the other coming from the acceleration needed 
to hold a steady position in the local gravitational field of the  galaxy.  Hence, she is out of equilibrium, in contact with two thermal baths at different temperatures, except where
$a=a_\Lambda$.

 By the quantum equivalence principle\cite{Smolin}, these have to enter the effective action with the same coefficient, $\alpha$, which is indeed the coefficient of the term induced by a real thermal bath.  But
we hypothesize that they enter the effective action with opposite signs.  There are several
motivations for this choice but, ultimately this is an hypothesis which is in need of
verification.  One justification is that a positive cosmological constant imposes a negative pressure, another is that deSitter spacetime has been argued to be unstable to quantum
fluctuations, while AdS has no such instability, due to its being dual to an ordinary 
QFT.  This means that the system in $AdS$, being dual to an ordinary $CFT$, must have a 
positive specific heat.

Since in both cases 
$T^2 \approx \Lambda$ either the deSitter case or the AdS case has to come into 
the effective potential with a destabilizing minus sign, relative  to ordinary thermal effects;
for the reasons stated it seems the former is more likely.

Alternatively we could take the view that this is part of the definition of the
dark matter.

We hence choose the signs so that the acceleration of the static observer acts like an ordinary temperature to restore symmetry breaking, while a cosmological constant acts against this tendency.  The thermal contribution to the non-relativistic potential energy is then,
to quadratic order,
\f
{\cal V}_{T}=  
\frac{\alpha}{m}  ( T^2_a - T^2_\Lambda )   |\Phi |^2 
\ff

The specific potential energy is then
\f
\bar{\cal V}_{T}= \frac{{\cal V}_{T}}{m}=
\frac{\alpha}{m^2}  ( T^2_a - T^2_\Lambda )   |\Phi |^2 
=\frac{\alpha D^2 }{4 \pi^2 c^2} (a^2 -a_\Lambda^2 ) |\Phi |^2 
\label{VT}
\ff
plus possible higher order terms needed to stabilize the potential.  
We note that the weak, non-relativistic quantum equivalence principle, as we stated it above is satisfied,
as the dependences of the mass and of $\hbar$ combined into a dependence on the
diffusion constant $D$.  This justifies the inclusion of these terms in the effective potential.

There is then a critical point where
\f
\alpha D^2 (a^2 -a_\Lambda^2 )  =0 
\ff

This occurs at a radius with an acceleration
\f
a_0 = a_\Lambda 
\ff
We hypothesize that near this critical point, the dynamics is invariant under scale
transformations, defined in  (\ref{scaling}) below and a phase transition takes place, as is singled by the
change of sign on the quadratic term of the effective potential.  
The quadratic term of the
effective potential becomes negative for $a < a_\Lambda$, signalling the spontaneous breaking of the $U(1)$ global symmetry of the $GP$ equation, allowing the emergence of a meaningful macroscopic
phase. 

In a typical galaxy the acceleration, $a$ experienced by the static observer, first rises rapidly near the centre, and after falls off as we move away  from the centre.  There is then
a radius $r=r_0$ at which $a$ falls to $a_0$.  There are then two regions.

\begin{itemize}

\item{} $ r < r_0$,  $ a >  a_0$ {\it The inner, symmetric or normal phase.}

\item{}$ r > r_0$, $ a <  a_0$ {\it The outer, broken symmetry or superfluid phase.}

In this region the dark matter condenses to a superfluid ring.  Note that there is an
outer boundary to this region, at some $r=r_2 > r_0$, whose 
physics we discuss below\cite{lsb}.

\end{itemize}


\subsection{The physics of the outer region}

The outer region extends to an $r=r_2 > r_0$ at which $a$ has fallen sufficiently below
$a_0$ that we are deep in the broken symmetry phase and the scale invariance associated
with the second order phase transition at $r_0$ no longer governs the physics.

In most galaxies, $r_2$ need only be a factor of $\approx 5 r_0 $ to match the rough flatness of the rotation curves observed.

Between $r_0$ and $r_2$ the physics of the dark matter condensate is governed by the two symmetries:
broken $U(1)$ invariance and scale invariance (\ref{scaling}).

The mass distribution of the dark matter then organizes itself so that scale invariance
is satisfied.  This results in (\ref{solution}), and (\ref{rho1}), which we saw above gives
a static solution to the GP equations.

 So we anticipate as we move out from $r_0$ to see a soft breaking of the scale invariance, which would be
measured by
\f
\frac{\partial v}{\partial \lambda} \approx | T_0 - T |^\alpha \approx | a_0 - a |^\alpha
\ff
where $\alpha$ is an unknown critical exponent.
This tells us that the orbital velocity gets a small $r$ dependence given by
\f
v=\bar{v}+ A \ln \frac{T}{T_0} \approx \bar{v}+ B \ln \frac{a}{a_0} =
\bar{v}- B \ln \frac{r}{r_0} 
\label{log}
\ff
for some  $B$.  We note that related observations are discussed in \cite{log}.

We would expect that for $r >> r_2 $ the velocity  curve returns asymptotically to its 
Newtonian behaviour, 
$v \approx \frac{\sqrt{G M_{tot}}}{r^{\frac{1}{2}}      }$

\section{Conclusions}

We take a new approach to the hypothesis that dark matter has a superfluid phase based
on the equivalence principle applied in the domain of many body quantum theory.  In this letter we sketched a simplified model, which serves to illustrate our hypotheses, but it will need to be developed if it is to compete with existing realistic models of galaxies.

This is based on several novel assumptions and observations.

\begin{itemize}

\item{}We see how the leading terms of the classical limit of the physics of a bose gas
self-interacting under its mutual gravitational attractions become independent of the
inertial and passive gravitational masses, of the particles.  In this case the physics is governed by a macroscopic wave function, $\Phi (r )$, whose dynamics is governed by the GP equation which coincides with the Newton-Schrodinger equation.

\item{} We hypothesize that the dark matter in a galaxy is subjected to two thermal
baths, the first due to the cosmological constant, at the deSitter temperature, $T_\lambda$;
the second the Unruh temperature due to the acceleration of a static observer in the gravitational field of the galaxy, $T_a$.  We posit that these come into the effective Hamiltonian with opposite signs.

\item{}Thus, at a radius $r_0$ at which  $T_a = T_\Lambda$, and hence 
$a = a_\Lambda = c^2 \sqrt{\Lambda }$, there is a scale invariant point and hence a second
order phase transition that signals the condensation of a Bose-Einstein condensate
for $r> r_0$.  The physics in this region is scale invariant, this means that the condensate
organizes its distribution, as a function of radius to achieve scale invariance.

\item{}  The physics near $r \approx r_0$ is scale invariant under $t \rightarrow \lambda t$  and $x  \rightarrow \lambda x$, which implies a flat rotation curve.  We would expect to see departures from scale invariance
as we move away from $r_0$, expressed as a logarithmic correction to the flat rotation curve,
as  in (\ref{log}).

\item{}The condensate behaves like a ring of super-currents, with a common, current velocity, which is rotational.  This seems to require coupling to a dark magnetic field.

\item{}  As a super-current, the condensate has  a well defined macro wavefunction and phase, $\phi $.  The fluid current is $J_a = \partial_a \phi + A_a$ which by scale invariance is
constant in the critical ring.


\item{}The stars also have a flat rotation curve in the critical region.   Even if they are much heavier than the dark matter particles, a star is still in the test mass approximation, so far as its motion in the galaxy is concerned. 

\item{}This simple  scenario explains the observed acceleration relation, the baryonic Tully Fisher relation and the success of the MOND
hypothesis in galaxies.  We note that while Milgrom long ago pointed out the importance of
scale invariance for MOND phenomenology\cite{MM-scale}, it is still impressive how much of that phenomenology, (i.e. flat rotation curves and baryonic Tully Fisher) is a consequence of scale
invariance alone.   Our new hypothesis is that this scale invariance is due to the dark matter
undergoing a second order phase transition in the region normally associated with MONDian behaviour.  


\item{}Our scenario provides a natural outer limit to the applicability of MONDian phenomenology.  

\end{itemize}

There remains much to be done to build this sketch of a scenario out into a detailed proposal.
Among the most urgent questions to investigate are.

\begin{enumerate}

\item{}Extend the two dimensional self-consistent solution we found to the semi-classical limit of the G-P (Newton-Schrodinger) equations to a fully three dimensional solution of that equation coupled
to the dark Maxwell equations.  

\item{}The dark magnetic field is trapped perpendicular to the ring, but will presumably
curl around the ring of dark super-current, making the galaxy into a dark-magnetic dipole.
Depending on the strength of this new interaction, this could give a halo-like component
to the dark matter density and contribute to the interactions between nearby galaxies.

\item{} Investigate a two fluid model, describing the interplay of a normal phase and superfluid phase.  This would also allow us to investigate the relative importance of halo and disk components of the dark matter.  Another interesting question to address in future work is whether the fact that the dark matter superfluid only condenses for $r>r_0$ addresses the cusp problem.

\item{}Our proposal that scale invariant behaviour is a consequence of a second order phase transition arising from a competition between the effects of Unruh and deSitter, horizon radiation needs more investigation.

\item{}MOND has been applied to, and made successful predictions for, other galaxy types including low surface brightness, ellipticals, dwarf ellipticals, dwarf-spheroidal satellites of the Mlkey Way and Andromeda; so we should understand if our proposal addresses these cases.

\item{}There are a number of astronomical systems in which the accelerations are near or below $a_0$, which include the Oort cloud, double and multiple star systems, the nearest one of which may be in the MOND regime, is alpha-Proxima.  A few exoplanets also appear to have accelerations near $a_0$\cite{inprogress}.  It would be very interesting to understand the physics of this regime, because the many body physics we described here may not be relevant.

\end{enumerate}

\section*{ACKNOWLEDGEMENTS}

We dedicate this paper to Robert Brout, who encouraged each of us to think about
how ideas from condensed matter physics might illuminate cosmological puzzles. 

We are grateful to Marina Cortes, Laurent Freidel, Sabine Hossenfelder, Andrew Liddle,
 Evan McDonough, Bavithra Naguleswaran,  Dam Son, David Spergel and Yigit Yargic 
for advice, encouragement, comments and correspondence.  We are especially
thankful to Mordehai Milgrom and Philip Phillips for close readings of a draft of  this paper.

This research was supported in part by Perimeter Institute for Theoretical Physics. Research at Perimeter Institute is supported by the Government of Canada through Industry Canada and by the Province of Ontario through the Ministry of Research and Innovation. This research was also partly supported by NSERC and FQXi and by a generous grant from the John Templeton Foundation.

\appendix

\section{Appendix: Different masses}

We can go a bit further, and ask whether we can extend our use of the non-relativistic quantum equivalence principle to the case where there are several species of dark matter particles, with different masses, $m_I$.  We note that, so long as $N$ is large, we can gain an approximation to the average bulk gravitational potential by replacing $m$ in (\ref{Sch1}) by
the average mass, $\bar{m}$, defined as
\f
\bar{m} =\frac{1}{N} \sum_I m_I
\ff
The masses have a distribution function, $\mu (m)$,  a possible form of which is a Gaussian
\f
\mu (m) \approx e^{-\frac{(m-\bar{m})^2}{2w^2}}
\ff

The key idea is that, by the equivalence principle, in general and in the non-relativistic quantum equivalence principle, the masses should be  irrelevant in the test particle
approximation, in the classical limit, even if they are different.  We might wonder whether  this  means that there is a regime of quantum gravitational physics where particles with different masses behave indistinguishability.
The particles would be quasi identical, due to the degeneracy of the classical limit imposed by the
equivalence principle. We call these {\it equivalent particles.}
We can hypothesize that in these cases the semiclassical wave function should be symmetric
under exchange of these equivalent particles.

If this is the case then a  consequence may be that in the limit $N \rightarrow \infty$,  the equivalent particles can condense into the ground state, even if they have different masses, and participate in the physics of the quantum fluid. 

This might be relevant for the stars in a disk galaxy.  When the dark matter mass is small enough, there are a large number, $P$, of dark matter particles for every star.  Adding the stars to the fluid barely changes the average mass.  While a star is much more massive than a 
dark matter particle, it is of negligible mass ($\approx 10^{-11}$), compared to the galaxy as a whole.

This might mean that the velocity field of the superfluid is shared by the stars. 

We can formulate this intuition as a conjecture:

Consider the gravitational potential per unit passive gravitational mass
\f
V(x) = - \sum_{J} \frac{G  m_J}{|x_J-x |}
\label{V}
\ff
We compare this to a different potential $V$, eq (\ref{V}) in which the masses
are each replaced by the averaged mass.
\f
\bar{V}(x,t) = - G \bar{m} \sum_J \frac{1}{|x-x_J |}
\label{barV}
\ff

\begin{itemize}

\item{} {\bf Conjecture A:} There exists configurations of masses in which the particles of different masses are {\it well mixed}, so that the difference between $V$ and $\bar{V}$ goes away as we take the 
limit $N \rightarrow \infty$.
\f
|V(x) - \bar{V}(x) | = 0
\ff

\end{itemize}

Dividing through now by $\bar{m}$ we reach the conclusion that
 {\it a quantum self-gravitating system in the limit $N\rightarrow \infty$ and $\hbar \rightarrow 0$ treats the bodies making up the system as identical.  }
 The leading order dynamics is then given by
\f
\imath  \bar{D} \frac{\partial \Psi}{\partial t} = \left \{ -  \frac{\bar{D}^2}{2} \sum_I \nabla_I^2 
- G \bar{m}  \sum_{I< J} \frac{1}{|x_J-x_I |}
\right \} \Psi
\label{SchD}
\ff
where we now define the averaged diffusion constant
 \f
\bar{D}= \frac{\hbar}{\bar{m}}
\label{barD}
\ff
Therefor, {\it a quantum self-gravitating system in the limit $N\rightarrow \infty$ and $\hbar \rightarrow 0$, and in the test-particle approximation, treats the bodies making up the system as identical, even when they have different masses.  }




\begin{thebibliography}{99}

\bibitem{Vera}V. C. Rubin, N. Thonnard, and W. K. Ford, ApJ, 225: L107 (1978);
V. C. Rubin, W. K. Ford, and N. Thonnard, ApJ, 238: 471 (1980).

\bibitem{TF}R. B. Tully and J. R. Fisher, Astron. Astrophys. 54, 661
(1977).

 \bibitem{Moti1}Milgrom proposes the BTFR be called instead the mass-asymptotic-speed-relation (MASR) to distinguish it from the original Tully-Fisher relation.
 
 
\bibitem{small-scatter}Stacy McGaugh, Federico Lelli, Jim Schombert,  {\it The Small Scatter of the Baryonic Tully-Fisher Relation}, The Astrophysical Journal Letters, Volume 816, Issue 1, article id. L14, 6 pp. (2016), arXiv:1512.04543.

\bibitem{McGaugh:2000sr} 
  S.~S.~McGaugh, J.~M.~Schombert, G.~D.~Bothun and W.~J.~G.~de Blok,
  ``The Baryonic Tully-Fisher relation,''
  Astrophys.\ J.\  {\bf 533}, L99 (2000)
  doi:10.1086/312628
  [astro-ph/0003001].
  
\bibitem{Milgrom} 
  M.~Milgrom,
  ``A Modification of the Newtonian dynamics as a possible alternative to the hidden mass hypothesis,''
  Astrophys.\ J.\  {\bf 270}, 365 (1983).
  doi:10.1086/161130
  
  
\bibitem{MS}Milgrom, Mordehai; Sanders, Robert H.,  {\it Rings and Shells of "Dark Matter" as MOND Artifacts}, arXiv:0709.2561,
The Astrophysical Journal, Volume 678, Issue 1, article id. 131-143, pp. (2008).

\bibitem{MOND-review}For reviews, see B. Famaey, and S. McGaugh, Liv. Rev. Rel. 15, 10
(2012); M. Milgrom, Scholarpedia 9(6), 31410 (2014).
  
  
\bibitem{MLS}Stacy McGaugh, Federico Lelli, Jim Schombert,  {\it The Radial Acceleration Relation in Rotationally Supported Galaxies},  arXiv:1609.05917v1; R.H. Sanders, Astron. Astrophys. Rev., 2, 1 (1990)

  
  
\bibitem{teves}J. D. Bekenstein, Phys. Rev. D 70, 083509 (2004) Erratum: [Phys. Rev. D 71,
069901 (2005)] doi:10.1103/PhysRevD.70.083509, 10.1103/PhysRevD.71.069901 [astroph/0403694].
                                                                                                                                                                                                 
\bibitem{moffat}J. W. Moffat, JCAP, 0505: 003 (2005); J. W. Moffat, JCAP, 0603: 004 (2006)
23M. Carmeli, Int. J. Theor. Phys, 37: 2621

\bibitem{dS}
Gibbons, Hawking. "Cosmological Event Horizons, Thermodynamics, and Particle Creation," Physical Review D, 15, 1977, pages 2738?2751
  
  \bibitem{Unruh}W.G. Unruh, {\it Notes on black hole evaporation}, Phys. Rev. D14, 870 (1976).
  
\bibitem{Milgrom2} Milgrom proposed in \cite{MM-related} that MOND might arise out of an interplay between the Unruh and deSitter temperature; this was also explored in
  Klinkhamer, F. R.; Kopp, M., {\it Entropic Gravity, Minimum Temperature, and Modified Newtonian Dynamics}, Modern Physics Letters A, Volume 26, Issue 37, pp. 2783-2791 (2011).
arXiv:1104.2022; 	Pazy, E.; Argaman, N., 
Quantum particle statistics on the holographic screen leads to modified Newtonian dynamics
as well as in \cite{Smolin}
 We believe that the scenario proposed here is novel.  

\bibitem{DL}S. Deser, Orit Levin,  {\it Accelerated Detectors and Temperature in (Anti) de Sitter Spaces}, Class.Quant.Grav.14:L163-L168,1997
DOI:	10.1088/0264-9381/14/9/003, arXiv:gr-qc/9706018;
Narnhofer H., Peter I. and Thirring W. 1996 Int. J. Mod. Phys. B 10 1507.

\bibitem{lsb}Note that in low surface brightness galaxies the whole disk will be in the
outer, superfluid phase.

\bibitem{MM-related}Mordehai Milgrom, {\it The Modified Dynamics as a Vacuum Effect}, Phys.Lett. A253 (1999) 273-279
DOI:	10.1016/S0375-9601(99)00077-8, arXiv:astro-ph/9805346;

\bibitem{MM-related2}Mordehai Milgrom{\it Dynamics with a Nonstandard Inertia-Acceleration Relation: An Alternative to Dark Matter in Galactic Systems},
arXiv:astro-ph/9303012, Annals of Physics, Volume 229, Issue 2, p. 384-415.

\bibitem{minic}Chiu Man Ho, Djordje Minic, Y. Jack Ng,  {\it Quantum Gravity and Dark Matter},
35. arXiv:1105.2916, Gen. Rel. Grav. 43 (2011) 2567-2573.
  
\bibitem{Silverman} 
  M.~P.~Silverman and R.~L.~Mallett,
  ``Dark matter as a cosmic Bose-Einstein condensate and possible superfluid,''
  Gen.\ Rel.\ Grav.\  {\bf 34}, 633 (2002).
  doi:10.1023/A:1015934027224
  
\bibitem{Sikivie} 
  P.~Sikivie and Q.~Yang,
  ``Bose-Einstein Condensation of Dark Matter Axions,''
  Phys.\ Rev.\ Lett.\  {\bf 103}, 111301 (2009)
  doi:10.1103/PhysRevLett.103.111301
  [arXiv:0901.1106 [hep-ph]].
  
\bibitem{Berezhiani:2015bqa} 
  L.~Berezhiani and J.~Khoury,
  Phys.\ Rev.\ D {\bf 92}, 103510 (2015)
  doi:10.1103/PhysRevD.92.103510
  [arXiv:1507.01019 [astro-ph.CO]].
  
  
\bibitem{Alexander1} 
  S.~Alexander and S.~Cormack,
  ``Gravitationally bound BCS state as dark matter,''
  JCAP {\bf 1704}, no. 04, 005 (2017)
  doi:10.1088/1475-7516/2017/04/005
  [arXiv:1607.08621 [astro-ph.CO]].
  
\bibitem{Alexander2} 
  S.~Alexander, E.~McDonough and D.~N.~Spergel,
  arXiv:1801.07255 [hep-th].
  
  
  
\bibitem{Smolin} 
  L.~Smolin,
 ``MOND as a regime of quantum gravity,''
  Phys.\ Rev.\ D {\bf 96}, no. 8, 083523 (2017)
  doi:10.1103/PhysRevD.96.083523
  [arXiv:1704.00780 [gr-qc]];  {\it Four principles for quantum gravity}, arXiv:1610.01968, contribution to Paddy@60, a book in honour of  Thanu Padmanabhan.
  
  
 \bibitem{MM-scale}M. Milgrom, Scale Invariance at low accelerations (aka MOND) and the dynamical anomalies in the Universe, arXiv:1605.07458v2;   MOND laws of galactic dynamics,
 Monthly Notices of the Royal Astronomical Society, Volume 437, Issue 3, p.2531-2541,
 (arXiv:1212.2568.

 
 
  
\bibitem{Reade} 
  J.~I.~Read, G.~Lake, O.~Agertz and V.~P.~Debattista,
  ``Thin, thick and dark discs in LCDM,''
  Mon.\ Not.\ Roy.\ Astron.\ Soc.\  {\bf 389}, 1041 (2008)
  doi:10.1111/j.1365-2966.2008.13643.x
  [arXiv:0803.2714 [astro-ph]].
  
  \bibitem{thick}Katelin Schutz, Tongyan Lin, Benjamin R. Safdi, Chih-Liang Wu,  
{\it Constraining a Thin Dark Matter Disk with Gaia}, 	arXiv:1711.03103.

  
  
\bibitem{Spergel} 
  D.~N.~Spergel and P.~J.~Steinhardt,
  ``Observational evidence for selfinteracting cold dark matter,''
  Phys.\ Rev.\ Lett.\  {\bf 84}, 3760 (2000)
  doi:10.1103/PhysRevLett.84.3760
  [astro-ph/9909386].
  
  
\bibitem{JiJi} 
  J.~Fan, A.~Katz, L.~Randall and M.~Reece,
  ``Dark-Disk Universe,''
  Phys.\ Rev.\ Lett.\  {\bf 110}, no. 21, 211302 (2013)
  doi:10.1103/PhysRevLett.110.211302
  [arXiv:1303.3271 [hep-ph]].
  
\bibitem{Saavas} 
  K.~Vattis and S.~M.~Koushiappas,
  ``Self-interacting dark matter constraints in a thick dark disk scenario,''
  arXiv:1801.06556 [astro-ph.GA].

\bibitem{log} S.S. McGaugh, W.J.G. de Blok, J.M. Schombert, R. Kuzio de Naray, J.H. Kim,
{\it The Rotation Velocity Attributable to Dark Matter at Intermediate Radii in Disk Galaxies}
Astrophys.J.659:149-161,2007
DOI:	10.1086/511807, arXiv:astro-ph/0612410

\bibitem{GP} E. P. Gross (1961). {\it Structure of a quantized vortex in boson systems}. Il Nuovo Cimento. 20 (3): 454?457. Bibcode:1961NCim...20..454G. doi:10.1007/BF02731494.
 L. P. Pitaevskii (1961). {\it Vortex lines in an imperfect Bose gas}. Sov. Phys. JETP. 13 (2): 451?454.

\bibitem{inprogress}
Bavithra Naguleswaran, Lee Smolin, and Robert Spekkens, {\it Astronomical regimes 
for MOND}, preprint in preparation.

\end{thebibliography}
\end{document}